\documentclass[11pt]{article}
\usepackage[margin=1in]{geometry}
\usepackage{times}
\usepackage{amsmath,amssymb}
\usepackage{graphicx}
\usepackage{booktabs}
\usepackage{array}
\usepackage{caption}
\usepackage[hidelinks]{hyperref}
\usepackage{authblk}
\usepackage[round,authoryear]{natbib}
\usepackage{titlesec}
\usepackage{enumitem}
\usepackage{xcolor}
\usepackage{microtype}
\titleformat{\section}{\normalfont\large\bfseries}{\thesection}{0.6em}{}
\titleformat{\subsection}{\normalfont\normalsize\bfseries}{\thesubsection}{0.6em}{}
\newcommand{\sym}[1]{\textsuperscript{#1}}

\title{\textbf{Platform Choice, Trust, and Privacy\\ in the Consumer AI Assistant Market}}
\author[ ]{Jennifer Zou\thanks{Profound. Contact: \texttt{jennifer@tryprofound.com}.}}
\date{July 2026}

\begin{document}
\maketitle

\begin{abstract}
\noindent We study how a representative sample of United States adult AI-assistant users ($n=1{,}999$; June 2026) choose among platforms, allocate tasks across them, evaluate provider trustworthiness, and value data-handling features. Estimates are weighted to the AI-user population using external adoption benchmarks. Four patterns emerge. The market is concentrated but internally differentiated: ChatGPT is the primary assistant for 58\% of users and Gemini for 25\%, yet smaller platforms hold defensible task niches---Claude captures a third of coding tasks despite a 7\% overall share. Task allocation is thus organized by platform far more than by user, and technical use falls steeply with age. Trust is earned through use rather than reputation: Claude is ranked most trustworthy in every head-to-head among users of both platforms, and shows by far the largest gap between how its users and non-users rate it. Finally, privacy concern is near-universal but action is gated by knowledge, not concern; in a choice experiment users pay most to keep humans---not models---out of their conversations (\$11.20/month), with valuations rising in task sensitivity.
\end{abstract}

\section{Introduction}
AI assistants are becoming increasingly integrated into the everyday: users route factual questions, work products, code, personal deliberations, and emotional disclosures through a small number of general-purpose conversational platforms. Unlike search engines, these systems answer directly rather than returning links, and repeated use accumulates provider-specific context---conversation history, saved preferences, uploaded files---that raises the cost of switching. The depth of disclosure and the stickiness of accumulated context make platform choice consequential both for users, whose sensitive information is at stake, and for the market, where early concentration may prove durable.

Yet the consumer side of this market is not well measured. Platform telemetry is proprietary and does not observe cross-platform behavior, privacy attitudes, or the reasons behind task-level choices. This paper reports a survey and embedded choice experiment designed to measure exactly these unobserved dimensions in a sample constructed to represent the US adult AI-using public. Leveraging this data, we present a reduced-form empirical analysis: descriptive market structure with proper uncertainty quantification, regression evidence on privacy behavior, and a conditional-logit valuation of data-handling attributes.

This paper makes four empirical contributions. First, we provide a significance-tested map of market structure: reporting awareness, past-month usage, and primary-platform shares for eight platforms with confidence intervals and testing of which rank orderings are statistically distinguishable. We find primary market share ranking is separated down to fifth place, while the awareness and usage funnels contain genuine ties. Second, and most distinctively, we measure choice at the task level rather than the user level. Because respondents report a primary platform separately for each of six task categories, we recover platform ``task signatures''---what each tool is actually used for---and the task-level substitution patterns (which platform users turn to when their primary is unavailable) that reveal where genuine competition occurs. This task-occasion granularity is absent from platform telemetry, and it shows that substitution is concentrated in specific tasks rather than spread evenly across a user's activity. Third, we measure trust both head-to-head, restricting to users who have experience with both platforms in a pair, and by experience, contrasting how users and non-users perceive each platform. Fourth, we estimate the dollar value users place on data privacy features through a discrete-choice experiment, and connect these valuations to a behavioral gap in which knowledge, not concern, predicts protective action. A methodological appendix documents the weighting scheme used to project the sample onto the AI-user population and additional robustness checks.

\section{Related Work}
This paper connects three literatures. The first is the privacy paradox---the recurring gap between stated privacy concern and protective behavior. \citet{acquisti2015} survey the economics of privacy and the frequent divergence between attitudes and actions, and \citet{athey2017} show experimentally that small frictions and incentives dominate stated preferences in determining whether individuals protect personal data. \citet{norberg2007} first documented the attitude--behavior gap directly. We contribute a decomposition specific to AI assistants: the gap is better explained by information---whether users know a platform's training policy---than by indifference, echoing evidence that disclosure and salience, rather than preference, drive privacy choices \citep{tucker2014}.

The second is the economics of switching costs and platform competition. \citet{klemperer1995} and \citet{farrell2007} establish how switching costs soften competition and entrench incumbents, and \citet{shapiro1999} describe lock-in in information-goods markets. A central question for AI assistants is whether observed single-homing reflects genuine quality sorting or behavioral lock-in; we measure functional reliance---tasks actually routed to more than one platform---as distinct from mere trial, and elicit task-level substitution, following the tradition of using diversion patterns to characterize competition \citep{conlon2020}.

The third is survey measurement of AI adoption. \citet{bick2024} use the Real-Time Population Survey to track generative-AI adoption in the United States and document its steep education and age gradients, and Pew Research \citep{pew2025} reports adoption by demographic group. These studies measure \emph{whether} people use AI; our contribution is to measure, within the user population, \emph{which} platform is used \emph{for which task}, and what users would pay to change how their data is handled---dimensions that adoption surveys and proprietary platform telemetry do not observe.

\section{Data and Methods}
\subsection{Sample and instrument}
We fielded a survey to US adults recruited through Prolific in June 2026, screening for awareness and past-month use of at least one AI assistant. The analysis sample is $n=1{,}999$ users (a small set of aware non-users is excluded from user-level estimates); median completion time was 11 minutes. The instrument elicited platform awareness and usage, per-task primary and secondary platform choices across six task categories, perceived task sensitivity, provider trust rankings, privacy attitudes and protective behavior, subscription tier, demographics, and a discrete-choice experiment over hypothetical plans. Because the frame is aware users by construction, all estimates describe the AI-using public, not the general adult population.

\subsection{Weighting to the AI-user population}
Survey samples rarely match their target population exactly, and post-stratification weighting is the standard remedy: respondents are reweighted so that the weighted sample reproduces known population margins (age, education, region, and so on), reducing bias from differential recruitment and nonresponse \citep{deville1992,bick2024}. When the reweighting margins are correct and the within-cell samples are not too thin, this removes composition bias at a modest cost in variance, summarized by the design effect. Online-panel studies of technology adoption routinely rake to Census margins in exactly this way \citep{bick2024,pew2025}.

The subtlety in our setting is the choice of target. The conventional target is the general adult population, but that is the wrong reference here: our sample is drawn from AI users by construction, so raking to general-population margins would reweight the sample toward the demographic profile of non-users and inject bias rather than remove it. Because adoption is far from uniform---it rises steeply with education and falls with age \citep{bick2024,pew2025}---the AI-user population differs systematically from the adult population, and the correct target is the former. 

We construct such targets by combining published adoption gradients with Census microdata. Let $N(a,e)$ be the 2024 American Community Survey (PUMS) adult population in age band $a$ and education band $e$. Pew Research reports AI-assistant adoption rates by age and by education; we reconcile these two marginal gradients into a coherent per-cell adoption surface $r(a,e)$ by iterative proportional fitting, so that the population-weighted marginals of $r$ reproduce both Pew gradients. The implied AI-user composition on any dimension $x$ is then
\begin{equation}
\pi_{\text{AI}}(x)\;\propto\;\sum_{a,e} N(a,e)\,r(a,e)\,\Pr(x\mid a,e),
\end{equation}
evaluated from the PUMS joint distribution. We rake the survey to these age and education targets. For industry---where general-population and AI-user distributions diverge sharply---we layer an additional margin: sector-level generative-AI adoption from the Real-Time Population Survey (RTPS, February 2026 wave) multiplied by each sector's share of the employed population, raked within the employed-with-industry stratum. The primary weight (age, education, industry; all measured gradients) has design effect $1.5$ (effective $n\approx1{,}300$). A lighter variant that balances only age and education has design effect $1.1$ (effective $n\approx1{,}800$); we use it as the default for user-level estimates and the industry-augmented weight for industry-conditional results.

We considered, but did not adopt, a further extension that infers income and employment adoption gradients by pushing the age--education adoption surface through the PUMS joint under a conditional-independence assumption. Because those gradients are not independently measured and the assumption is strong, we relegate the resulting weights to a robustness comparison (Appendix~\ref{app:robust}); headline estimates do not depend on them.

\subsection{Estimation and inference}
Point estimates are H\'ajek ratio means, $\hat{\theta}=\sum_i w_i y_i/\sum_i w_i$. Standard errors use the linearization (sandwich) estimator for weighted ratio means,
\begin{equation}
\widehat{\operatorname{Var}}(\hat\theta)=\frac{n}{n-1}\,\frac{\sum_i w_i^2 (y_i-\hat\theta)^2}{\left(\sum_i w_i\right)^2},
\end{equation}
which accounts for the variance inflation from unequal weights. Reported intervals are the default 95\% ($z=1.96$). Comparisons of two mutually exclusive shares within the same respondents (e.g., one platform's primary share versus another's) use the within-respondent difference indicator $y_i=\mathbf{1}\{A\}-\mathbf{1}\{B\}$, whose weighted mean is the share difference and whose linearization variance absorbs the induced negative covariance; comparisons across independent groups use $\mathrm{SE}=\sqrt{\mathrm{SE}_A^2+\mathrm{SE}_B^2}$; paired task comparisons restrict to respondents present in both tasks. We test many hypotheses; the large-margin findings (market structure, task signatures, the privacy-action gap, WTP) are far from the threshold, while claims we mark ``borderline'' sit near $z=2.0$ and warrant the usual multiple-comparison caution.

\subsection{The discrete-choice experiment}
Each respondent completed six choice tasks, selecting between two hypothetical AI plans (or neither) that varied on monthly price and three data-handling attributes: whether conversations are reviewed by humans, whether they are used to train the model, and whether answers contain sponsored content. Task vignettes were drawn from the respondent's own task set to vary sensitivity. We estimate a conditional logit on plan choices (the outside option excluded from estimation) and convert attribute coefficients to dollar values by the ratio to the price coefficient. Sensitivity and concern interactions are estimated by splitting on the vignette's elicited sensitivity and on the respondent's self-reported data-use concern (measured 1-5).

\section{Market Structure}
\subsection{Awareness, usage, and primary choice}
Table~\ref{tab:shares} reports the three-stage funnel with 95\% intervals. ChatGPT and Gemini are near-universally known (98.9\% and 95.8\%); a middle tier of Claude, Copilot, and Meta~AI clusters around 77--79\%; Grok is lower; Perplexity and DeepSeek are recognized by fewer than half. Past-month usage compresses the market into three tiers: ChatGPT (85.7\%), Gemini (77.4\%), and a third tier led by a Copilot--Claude pair around one third. Primary platform choice\footnote{Formally, we define a respondent's primary platform as the one through which they route the majority of their tasks.} is the most concentrated and the most cleanly ordered.

\begin{table}[t]
\centering
\caption{Platform funnel: awareness, past-month usage, and primary share (weighted \%, 95\% CI half-widths).}
\label{tab:shares}
\small
\begin{tabular}{lccc}
\toprule
Platform & Aware & Used & Primary \\
\midrule
ChatGPT & $98.9\pm0.5$ & $85.7\pm1.7$ & $58.2\pm2.3$ \\
Gemini & $95.8\pm1.0$ & $77.4\pm1.9$ & $25.4\pm2.1$ \\
Claude & $79.4\pm1.9$ & $34.1\pm2.2$ & $7.0\pm1.2$ \\
Copilot & $78.7\pm1.9$ & $35.0\pm2.2$ & $4.1\pm1.0$ \\
Meta AI & $77.4\pm2.0$ & $20.5\pm1.9$ & $0.6\pm0.4$ \\
Grok & $68.5\pm2.2$ & $19.9\pm1.9$ & $1.8\pm0.6$ \\
Perplexity & $46.8\pm2.3$ & $10.1\pm1.4$ & $0.5\pm0.3$ \\
DeepSeek & $44.5\pm2.3$ & $9.7\pm1.4$ & $0.6\pm0.4$ \\
\bottomrule
\end{tabular}
\end{table}

For primary share, every adjacent rank down through fifth place is significantly separated: ChatGPT $>$ Gemini ($z=16.1$), Gemini $>$ Claude ($z=14.3$), Claude $>$ Copilot ($z=3.6$), and Copilot $>$ Grok ($z=3.8$). Only the bottom three (Meta~AI, DeepSeek, and Perplexity) are mutually indistinguishable; each has an upper 95\% bound at or below roughly 1\%, so we can state that they individually hold at most about 1\% of the primary usage market. In the awareness and usage funnels, by contrast, the Claude/Copilot/Meta~AI cluster (awareness) and the Copilot/Claude and Perplexity/DeepSeek pairs (usage) are statistical ties. 

Concentration, moreover, is not merely an artifact of reach. ChatGPT's usage lead over Gemini (85.7 vs 77.4) is modest relative to its primary-share lead (58.2 vs 25.4): conditional on using ChatGPT, users disproportionately make it their main tool. The primary-share gap therefore exceeds what reach alone would predict.

\subsection{Monetization and the consumer--enterprise divide}
How platforms convert users into paying customers reveals much about their market position. Table~\ref{tab:convert} reports paid conversion among each platform's users, split by whether the subscription is personally purchased or provided by an employer or school. Two facts stand out. First, Claude has the highest overall paid conversion of any platform---34\% of its users pay, edging out ChatGPT (32\%) and well ahead of Gemini (18\%)---despite Claude's small primary-user base. A platform can therefore be a minor player in share while monetizing its users especially well, a pattern consistent with Claude attracting deliberate, high-intent users (Section~\ref{sec:multihoming}). Second, the channel of payment cleanly separates a consumer from an enterprise orientation. Among ChatGPT's paying users, roughly four in five (79\%) pay out of pocket rather than through an employer; the same is true of Grok (87\%) and Perplexity (70\%). Copilot is the mirror image: only 28\% of its paid plans are personal, the rest employer- or school-provided, marking it as an enterprise-distributed product rather than a consumer purchase.

\begin{table}[t]
\centering
\caption{Paid conversion among each platform's users (weighted \%). ``Personal'' and ``work/school'' are the shares of \emph{users} on each plan type; ``\% of paid personal'' is the share of paying users who pay out of pocket.}
\label{tab:convert}
\small
\begin{tabular}{lcccc}
\toprule
Platform & Personal-paid & Work/school-paid & Any paid & \% of paid personal \\
\midrule
ChatGPT & 25 & 7 & 32 & 79 \\
Claude & 23 & 11 & 34 & 67 \\
Copilot & 8 & 20 & 28 & 28 \\
Gemini & 12 & 6 & 18 & 66 \\
Grok & 18 & 3 & 21 & 87 \\
Perplexity & 21 & 9 & 31 & 70 \\
DeepSeek & 15 & 5 & 20 & 76 \\
\bottomrule
\end{tabular}
\end{table}

\section{User Bases and Multihoming}\label{sec:multihoming}

\subsection{Distinct audiences}
\begin{figure}[t]
\centering
\includegraphics[width=0.82\textwidth]{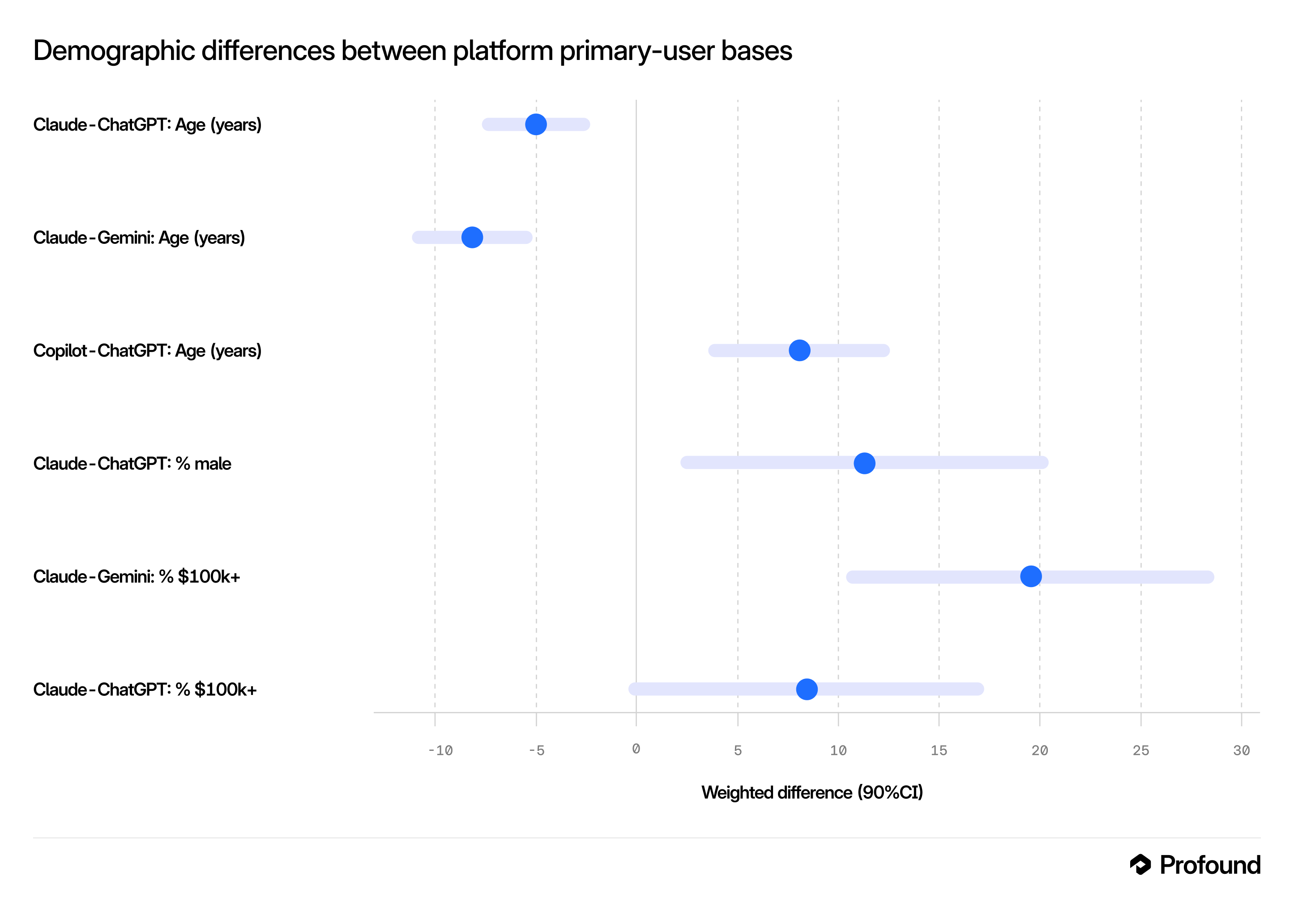}
\caption{Differences between platform primary-user bases (weighted, 95\% CI). Each row is a pairwise contrast; intervals excluding zero are significant at 5\%. Positive values indicate the first platform's users are older, more male, or higher-income than the second's. The Claude--ChatGPT income contrast is the one gap whose interval crosses zero.}
\label{fig:forest}
\end{figure}

The platforms sort their users along two key demographic axes---age and affluence---and Figure~\ref{fig:forest} plots the pairwise gaps with confidence intervals. Claude anchors one pole: its primary users are the youngest, most male, and highest-income of the major platforms. Copilot and Gemini anchor the other on age, drawing users roughly five to nine years older than ChatGPT's, while ChatGPT sits near the middle of every dimension, consistent with its role as the mass-market default. The age gaps are the most pronounced and precisely estimated---Claude's users average 33 against ChatGPT's 38 and Gemini's 42---and the income gap between Claude and Gemini is large (a 20-point difference in the share earning over \$100k). Most of these contrasts clear conventional significance, as the intervals in Figure~\ref{fig:forest} show; the marginal case is income between Claude and ChatGPT, where Claude's base has a visibly longer upper tail (more \$100k+ earners) but the two are statistically indistinguishable on the income scale as a whole.

A crucial distinction separates the \emph{profile} of a platform's users from the \emph{targeting} implication. Although Claude's user base skews young, young people do not disproportionately use Claude: among 18--24 year-olds, ChatGPT is the primary platform far more than Claude ($z=10.3$) and Claude's primary share is actually \emph{lower} than Gemini's ($z=3.2$). Because ChatGPT is so much larger, it dominates the primary slot in every age band. The correct reading is ``Claude's users are young,'' not ``the young use Claude.''

\subsection{Trial versus reliance}
Casual trial of multiple assistants is common; genuine reliance on more than one is less so. Four in five users have tried a second assistant, but only about 43\% intentionally route different tasks to more than one platform\footnote{We ask respondents to identify their primary platform for each completed task category; intentional routing requires that at least 2 distinct primary platforms were chosen.}. Among those who do, Claude's primary users are the most likely to multihome (51\%) and Gemini's the least (38\%), consistent with Claude attracting deliberate multi-tool users and Gemini serving as a settled default. Reported stickiness echoes this: Claude's primary users report the highest platform match quality (4.0/5) and the weakest habit-driven inertia (3.3) as reasons for platform choice, while Gemini's report the lowest switching cost (2.9).

\section{Task Allocation}
\begin{figure}[t]
\centering
\includegraphics[width=0.72\textwidth]{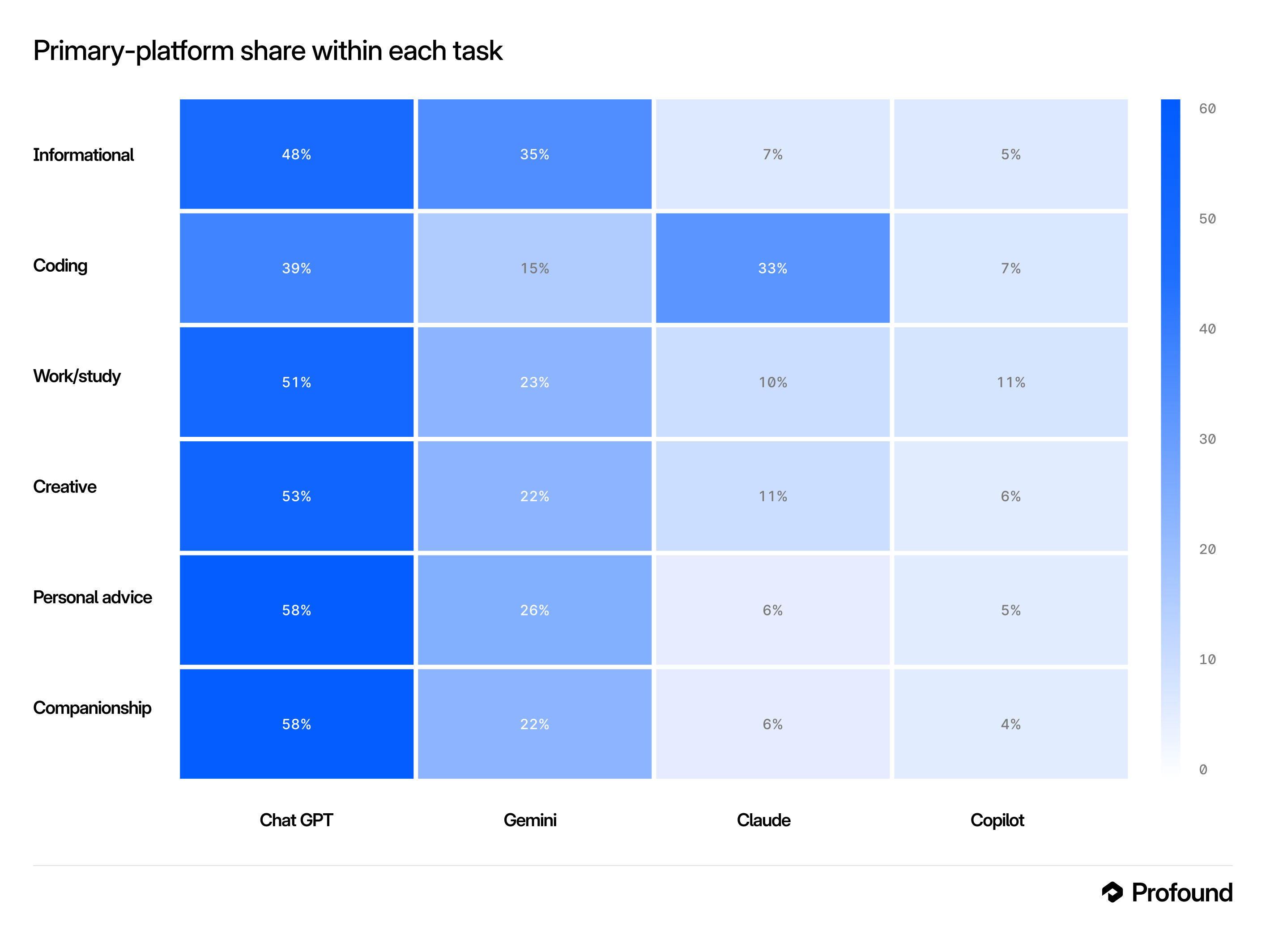}
\caption{Primary-platform share within each task category (weighted \%). Claude's coding concentration and Copilot's work tilt are visible against ChatGPT/Gemini generalism.}
\label{fig:heat}
\end{figure}

What people use AI for is differentiated more by platform than by person (Figure~\ref{fig:heat}). Within coding tasks, Claude holds 33\% primary share---against 7\% overall---nearly matching ChatGPT (39\%) and far exceeding Gemini (15\%). Claude is used significantly more for coding than for informational queries ($z=10.9$) or companionship ($z=5.9$). Copilot roughly doubles its share in work tasks compared to personal ones (such as companionship ($z=5.0$) or personal advice ($z=6.4$)). ChatGPT and Gemini are generalists tilted toward informational and everyday use; both are used more for informational queries than coding tasks ($z=3.8$ and $8.1$).

\begin{figure}[t]
\centering
\includegraphics[width=0.72\textwidth]{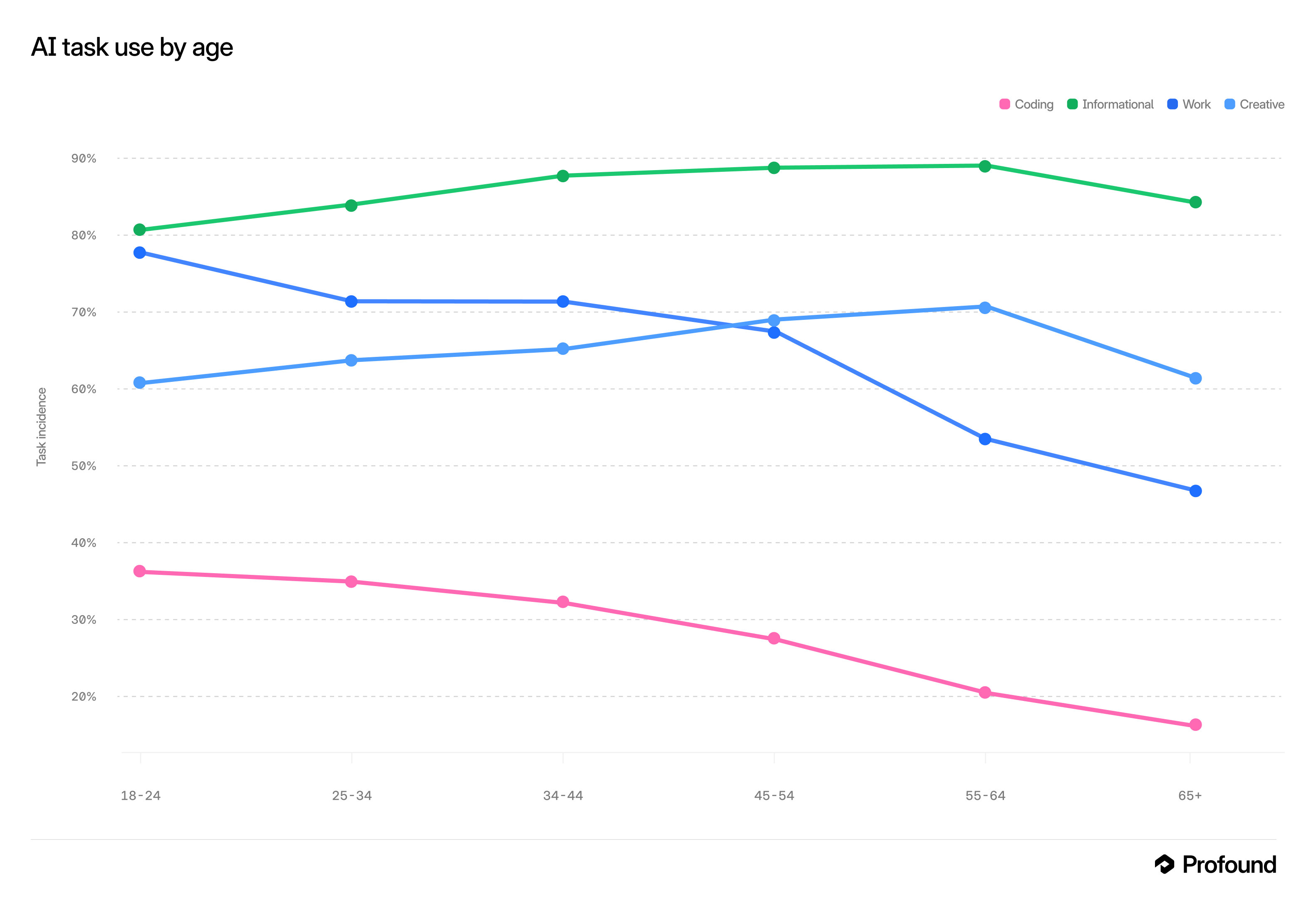}
\caption{AI task incidence by age band (weighted). Technical and work use decline with age; informational use rises.}
\label{fig:taskage}
\end{figure}

These signatures have sharp demographic gradients (Figure~\ref{fig:taskage}). Technical use falls steeply with age: coding incidence declines $-4.1$ percentage points per decade ($z=5.3$) and work/study $-5.7$ ($z=6.5$), while informational use rises modestly ($+1.5$/decade, $z=2.2$); creative, personal, and companionship use show no significant age trend. Task use also tracks industry. Coding is over-indexed in Information (60\% vs 29\% elsewhere, $z=7.3$) and Professional/scientific/technical services (57\%, $z=7.4$), and under-indexed in Other services, Retail, Arts, and Transportation. AI-for-work incidence is highest in finance, professional/technical services, information, and healthcare (all $\approx$80--84\%).

\section{Trust}
Trust is central to a market built on disclosure: users route sensitive information through these systems, so which provider they believe will handle it responsibly shapes both platform choice and how much they are willing to share. A first-order fact is that \emph{generalized} trust does not discriminate among the leaders---35--37\% of each major platform's users express trust in AI broadly. If trust mattered only in this diffuse sense, it would be a wash across platforms. The informative variation is instead relational and comparative: which provider a user ranks above another when forced to choose, and how that ranking depends on whether the user has firsthand experience.

We measure this two ways. The first is head-to-head. For each pair of platforms, we restrict to respondents who rank both and ask which is placed higher. This yields a clean, transitive ordering led by Claude (Table~\ref{tab:h2h}): Claude is ranked above ChatGPT (59--41) and Gemini (66--34), ChatGPT above Gemini (60--40) and Copilot (61--39), and Gemini above Copilot (57--43). It's important to emphasize that transitivity is not mechanical; rather, respondents are expressing a stable latent trust ordering. The ``ranked both'' column reports the effective base for each comparison.

\begin{table}[t]
\centering
\caption{Head-to-head trust rankings among respondents who ranked both platforms (weighted).}
\label{tab:h2h}
\small
\begin{tabular}{lcc}
\toprule
Comparison & Ranked both & Result \\
\midrule
Claude vs.\ ChatGPT & 592 & Claude higher, 59--41 \\
Claude vs.\ Gemini & 535 & Claude higher, 66--34 \\
ChatGPT vs.\ Gemini & 1{,}081 & ChatGPT higher, 60--40 \\
ChatGPT vs.\ Copilot & 556 & ChatGPT higher, 61--39 \\
Gemini vs.\ Copilot & 574 & Gemini higher, 57--43 \\
\bottomrule
\end{tabular}
\end{table}

\begin{figure}[t]
\centering
\includegraphics[width=0.72\textwidth]{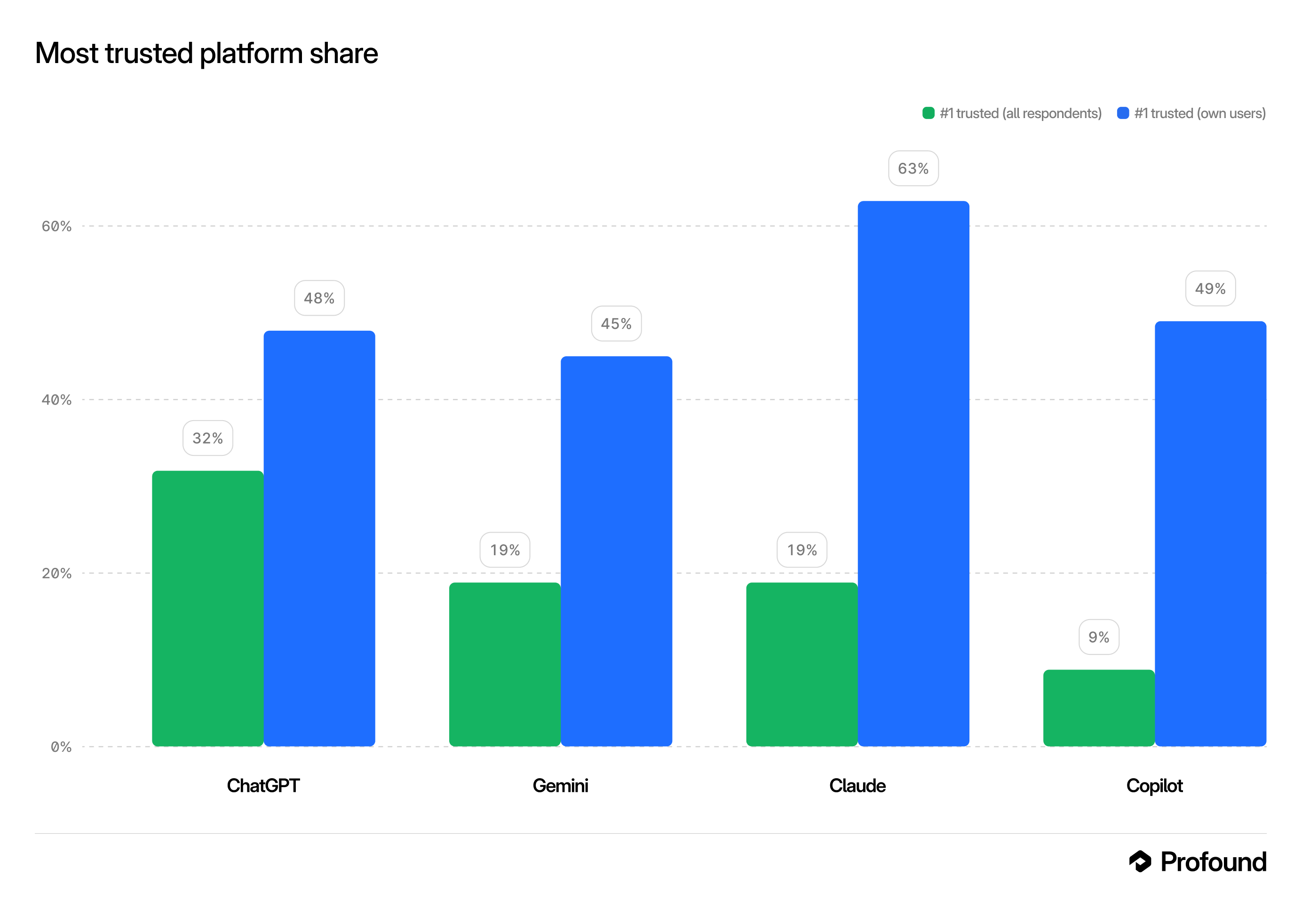}
\caption{First-choice trust: share naming each platform \#1 among all respondents versus among the platform's own users.}
\label{fig:trust}
\end{figure}

The second measurement investigates where trust comes from, and here one of the most striking findings of the paper emerges: for the challenger platforms, trust has to be earned through use rather than inherited from reputation (Figure~\ref{fig:trust}). ChatGPT and Gemini are already trusted by people who have never used them---54\% and 60\% of aware non-users rank them among their three most trusted platforms---plausibly riding the brand recognition of OpenAI and Google. Claude is different. Only 41\% of those aware of it but not using it rank it top-three, but that figure rises to 76\% among its actual users: a 35-point experiential lift, nearly 1.5 times ChatGPT's (24 points) and the largest of any platform. Put differently, Claude's reputation understates how its actual users regard it, while the incumbents' reputations roughly match or exceed their users' experience.


This gap between reputational and experiential trust has a general competitive implication that applies across platforms. Where trust tracks brand awareness, as it does for the incumbents, the measured level is close to a ceiling: most of the population that will hold a favorable view already does, without direct experience. Where the gap between users and non-users is large, as for Claude, awareness-based metrics understate how the platform is regarded by those who have used it, and that gap can close only through adoption rather than reputation. The two configurations imply different growth constraints, brand perception versus trial, though which matters more for long-run market outcomes depends on adoption dynamics we do not observe. The head-to-head results are consistent with this reading: among respondents who have used both platforms, the ordering favors the challenger (Claude) in all comparisons.

\section{The Privacy-Action Gap}
\begin{figure}[t]
\centering
\includegraphics[width=0.68\textwidth]{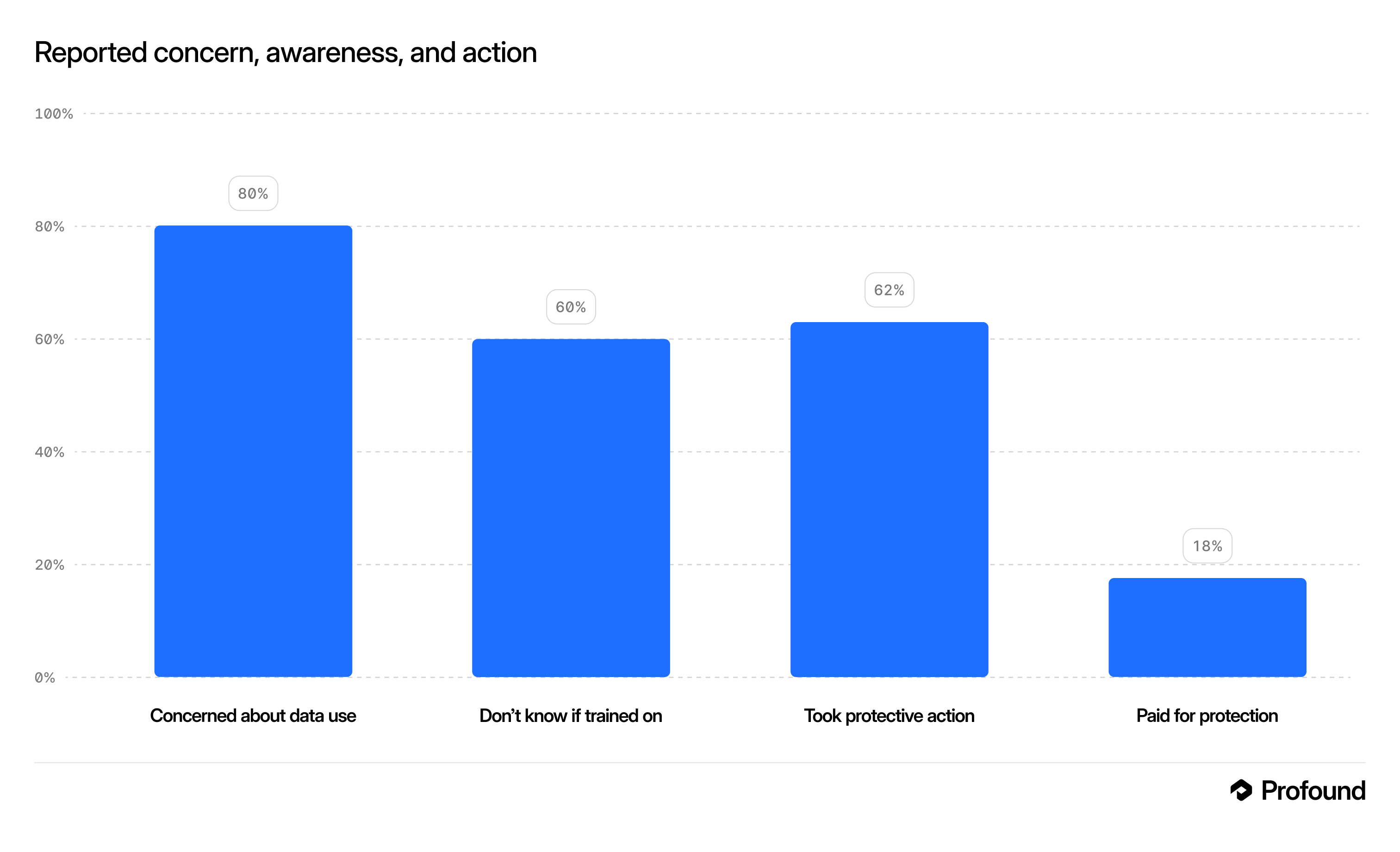}
\caption{The privacy-action gap: near-universal concern, widespread ignorance of training practices, and limited protective action.}
\label{fig:paradox}
\end{figure}

More than 80\% of users report that they are concerned about how their conversation data is used, yet protective behavior lags: roughly 60\% do not know whether their assistant trains on their conversations, and only 18\% have ever paid for a plan with better privacy protection (Figure~\ref{fig:paradox}). Concern is flat across demographic groups; what varies is knowledge and follow-through.

Table~\ref{tab:ols} reports a weighted OLS regression of the count of protective tools a user has adopted (temporary chat, history deletion, memory controls, signed-out use, and a privacy-motivated paid plan) on privacy attitudes, usage intensity, and demographics, with heteroskedasticity-robust standard errors. The specification is deliberately inclusive so that each attitude enters conditional on the others. The dominant predictor is not concern but \emph{policy literacy}: knowing whether one's assistant trains on conversations is associated with $0.34$ additional tools ($t=6.1$), the largest effect in the model and larger than a full point of concern on the five-point scale. Concern about data use matters ($0.17$ per point, $t=5.8$), as do stated willingness to pay, task sensitivity, and breadth of use, but their coefficients are half to a third the size of literacy's. At the user level, higher education and lower income are associated with more protective actions; engagement with these tools declines with user age. 

\begin{table}[t]
\centering
\caption{Selected determinants of protective behavior. Weighted OLS; dependent variable is the count of data-protection tools used (0--5). Heteroskedasticity-robust standard errors in parentheses. \sym{*}, \sym{**}, \sym{***} denote significance at 10\%, 5\%, and 1\%.}
\label{tab:ols}
\small
\begin{tabular}{lc}
\toprule
& Protective tools used \\
\midrule
Knows training policy & $0.341^{***}$ \\
 & $(0.056)$ \\
Employer restricts AI use & $0.170^{***}$ \\
 & $(0.061)$ \\
Concern about data use & $0.167^{***}$ \\
 & $(0.029)$ \\
Avg.\ task sensitivity & $0.117^{***}$ \\
 & $(0.027)$ \\
Platforms used & $0.120^{***}$ \\
 & $(0.020)$ \\
Education & $0.078^{***}$ \\
 & $(0.028)$ \\
Income & $-0.054^{***}$ \\
 & $(0.018)$ \\
Age & $-0.008^{***}$ \\
 & $(0.002)$ \\
Constant & $-0.086$ \\
 & $(0.200)$ \\
\midrule
Observations & 1{,}954 \\
$R^2$ & 0.182 \\
\bottomrule
\end{tabular}
\end{table}

The pattern reframes the privacy paradox as an information problem: literacy in training practices halves from the youngest to the oldest users while concern is flat, and protective behavior tracks literacy, not concern. The model explains a modest share of the variation ($R^2=0.18$), which is itself the point---protective behavior is only loosely tied to any single attitude and there still exists a large gap between concern and action.

\section{The Value of Privacy Features}
To value privacy features in money, we use the discrete-choice experiment described in Section~3.4. Each respondent saw six choices between two hypothetical assistant plans (plus an outside option), and the plans varied independently in monthly price and three privacy features: whether conversations are read by human reviewers, whether they are used to train the model, and whether answers contain sponsored content. Because the features and price were randomized independently, their effects on choice are separately identified.

We model the choice with a conditional logit. Respondent $i$ derives utility from plan $j$ according to:
\begin{align*}          
U_{ij}=\beta_{\text{review}}\,\mathrm{Review}_{j}+\beta_{\text{train}}\,\mathrm{Train}_{j}+\beta_{\text{ads}}\,\mathrm{Ads}_{j}+\beta_{\text{price}}\,\mathrm{Price}_{j}+\varepsilon_{ij}
\end{align*} where $\varepsilon_{ij}$ are i.i.d.\ extreme-value shocks. Let $V$ be the non-stochastic part of utility. The probability of choosing $j$, then, can be expressed as $\exp(V_{ij}) / \sum_k \exp(V_{ik})$. The coefficient on each feature gives its effect on choice utility; dividing a feature coefficient by the price coefficient converts it to a dollar value---the monthly price change that would offset the feature, i.e.\ willingness to pay to avoid it. The outside option is excluded from estimation so that valuations describe trade-offs among people who would choose some plan.

The estimated valuations reveal a steep hierarchy (Table~\ref{tab:wtp}). The feature users pay most to avoid is human review of their conversations, worth \$11.20/month---nearly four times what they pay to avoid their data being used for training (\$2.97), with avoiding sponsored answers in between (\$6.46). Users are far more troubled by the prospect of a person---an employer, a contractor, a government requester---reading their words than by a model ingesting them, even though public and regulatory attention centers on training. The feature that dominates policy debate is actually the one users value least of the three.

\begin{table}[t]
\centering
\caption{Attribute valuations from the conditional logit. ``Decision share'' is the percentage of choices in which respondents named the factor as the primary reason for their choice.}
\label{tab:wtp}
\small
\begin{tabular}{lccc}
\toprule
Attribute & Worth (\$/mo) & Choice prob.\ effect & Decision share \\
\midrule
No human review & \$11.20 & $+11.6$ pp & 29\% \\
No sponsored answers & \$6.46 & $+6.7$ pp & 16\% \\
No training on conversations & \$2.97 & $+3.1$ pp & 16\% \\
Price & (reference) & $-1$ pp per \$1 & 38\% \\
\bottomrule
\end{tabular}
\end{table}

\begin{figure}[t]
\centering
\includegraphics[width=0.7\textwidth]{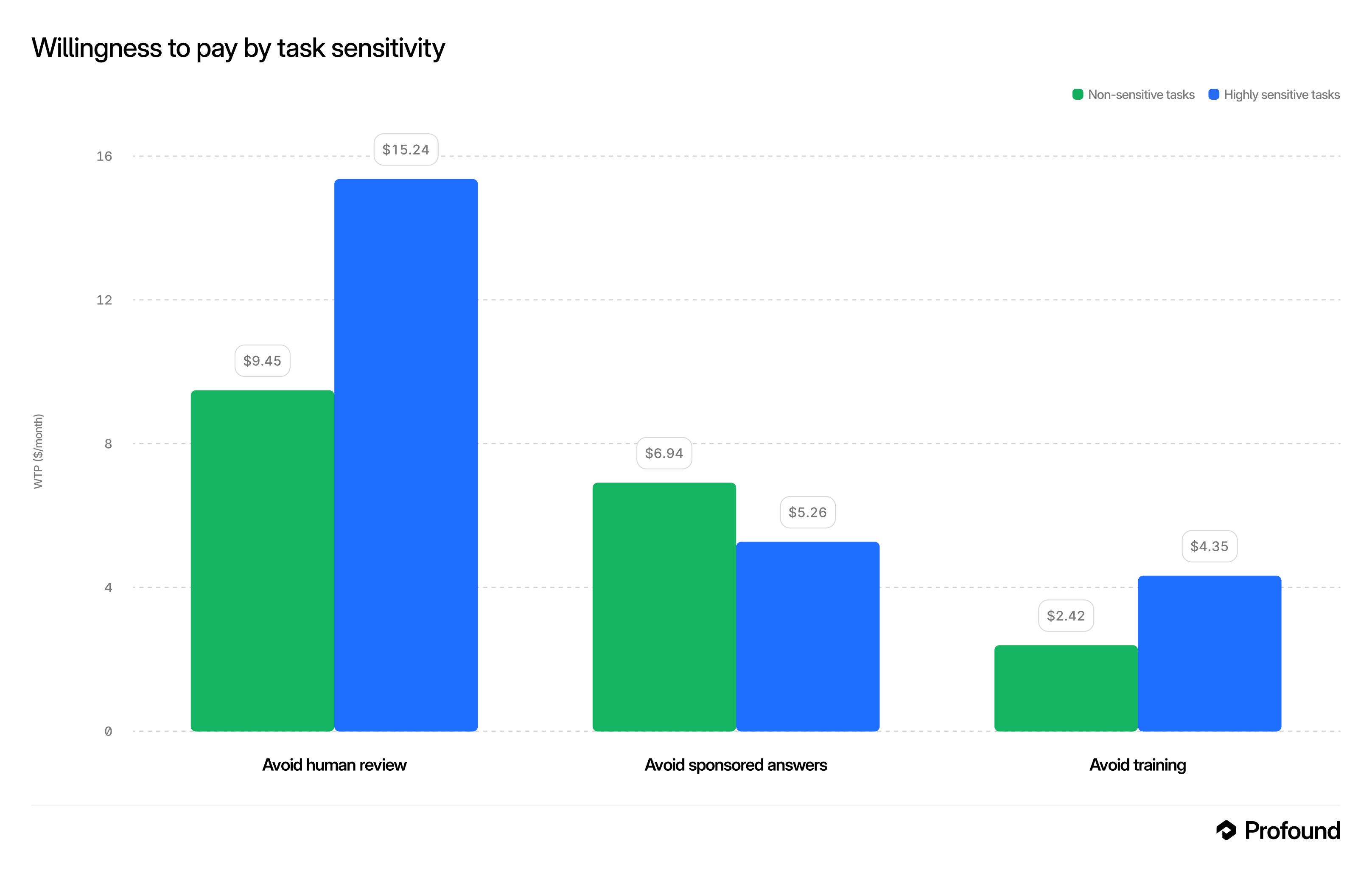}
\caption{Willingness to pay for each guarantee by task sensitivity. Human-review and training aversion rise with sensitivity; ad-aversion does not.}
\label{fig:wtpsens}
\end{figure}

Valuations also scale with stakes (Figure~\ref{fig:wtpsens}). Moving from non-sensitive to highly sensitive tasks, the value of avoiding human review rises from \$9.45 to \$15.24 and avoiding training from \$2.42 to \$4.35, while ad-aversion remains flat. The same pattern holds with user concern: among the most concerned users, avoiding human review exceeds \$15/month and avoiding training approaches \$7. Price nonetheless plays an important role in every choice---cited as the primary reason behind 38\% of selections---so privacy features can differentiate a plan but do not substitute for affordability.

\section{Discussion}
Three themes emerge from the results. First, the market is concentrated but differentiated. ChatGPT and Gemini dominate reach and primary choice, yet the smaller platforms are not simply scaled-down versions of the leaders: they hold defensible task niches (Claude in coding, Copilot in work), attract distinct audiences, and---in Claude's case---convert and retain users at rates the leaders do not match. Competition in this market is unlikely to play out as a single contest over aggregate share; it is better understood as a set of task-specific and segment-specific contests, most of which ChatGPT currently leads but some of which it does not.


Second, trust is organized differently from the rest of the market. The demographic and usage metrics that describe platform choice---reach, primary share, the age and income profiles of each user base---are all strongly correlated and led throughout by the incumbents. Trust is the one dimension that does not line up with reach; specifically: how much a platform's standing depends on firsthand experience. For the incumbents, trust is boosted by broad brand awareness and sits near the ceiling; for challenger platforms, the experiential gap between users and non-users is wide, so awareness-based measures may understate their standing. This implies that a platform's position on trust cannot be read off its size. Trust-based competition is therefore bounded less by reputation than by trial, a constraint invisible to the reach and share metrics that otherwise define the market.

Third, privacy behavior is an information problem, not a preference problem. Concern is near-universal and weakly predictive of action; knowledge of data practices is the strongest operative lever, and the feature users most want to control---human access to their conversations---is not the one that dominates public debate. Together these suggest that disclosure and defaults are what would move protective behavior, and that providers competing on privacy might do better to emphasize human-access guarantees than training policies.

There are limitations to the scope of these results. The frame is aware US adult users recruited through an online panel, so estimates describe the AI-using public rather than all adults; adoption-level questions must be answered from external benchmarks rather than from this sample. Industry results cover only employed respondents who disclosed a sector, which may introduce bias through selection. The choice experiment is stated-preference, so its dollar figures are relative valuations calibrated to a randomized price rather than revealed market prices. The smaller platforms have primary-user bases too thin to profile individually or to resolve platform-by-industry preferences, which is why per-sector platform comparisons focus on ChatGPT versus Gemini. Finally, we test many hypotheses; the large-margin findings are robust, but a handful of claims sit near the 5\% threshold and are flagged as such in the text.

\section{Conclusion}
Using a representative sample of US adults who use AI assistants, we have documented a market that is top-heavy in aggregate but internally differentiated: a cleanly ordered primary-platform hierarchy, platform-specific task niches, audiences that differ systematically by age and income, monetization models that separate consumer from enterprise products, and provider trust that is both earned through use and inherited from reputation. On privacy, concern is near-universal but action is gated by knowledge, and users will pay most to keep humans---not models---out of their conversations, with valuations that rise in the sensitivity of the task.

Several questions follow directly from these findings and are natural next steps. The task-level substitution patterns we measure invite a formal model of how users allocate occasions across platforms and how much switching is driven by task fit versus habit---a question our reduced-form estimates motivate but do not resolve. The experiential trust gap suggests a causal question our cross-sectional design cannot answer: does usage raise trust, or do trusting users self-select into adoption? A trial-based or longitudinal design could separate the two. The stated-preference valuations invite validation against revealed behavior, for instance by observing responses to actual privacy-feature launches or price changes. And because adoption is still rising and broadening, the demographic and task gradients we report are best read as a snapshot of a moving distribution; tracking them over time would show whether early niches (Claude in coding, the youth skew in technical use) harden into durable structure or erode as the market matures.

\clearpage
\appendix
\section*{\large Appendix}
\addcontentsline{toc}{section}{Appendix}
\section{Robustness and Additional Cuts}
\label{app:robust}
\subsection*{A.1\quad Weighting variants and their agreement}
We constructed several weighting schemes and use two for headline results: an age--education AI-user weight for user-level estimates and an age--education--industry weight for industry-conditional estimates. Table~\ref{tab:weights} reports, for each scheme, its design effect and effective sample size alongside four representative estimands and a summary agreement metric---the mean absolute deviation (MAD) of the scheme's estimates from our default across those estimands.

Two points follow. First, the cost of weighting is modest for the schemes we use: the default age--education weight has design effect $1.13$ (effective $n\approx1{,}800$), and even the industry-augmented weight retains an effective $n\approx1{,}320$. Second, and more importantly for interpretation, \emph{our headline estimates are insensitive to the weighting choice among defensible schemes but sensitive to the target-population error we caution against}. The default weight differs from the unweighted sample by only $0.6$ percentage points on average across the four estimands, and from the industry-augmented weight by $0.7$; the inferred-income variant we set aside differs by $1.6$. By contrast, general-population raking---the inappropriate target---moves estimates by $3.6$ points on average, concentrated in the ChatGPT and Gemini shares (it pulls ChatGPT primary share from 58\% down to 52\% and Gemini up from 25\% to 31\%). The methodological lesson is that in an aware-user sample, \emph{which} population one weights to matters far more than the mechanics of weighting: the schemes that target AI users agree closely with one another and with the raw data, while targeting the general population introduces the largest movement of any choice we examine.

\begin{table}[h]
\centering
\caption{Weighting schemes: design effect, effective $n$, four representative estimands (weighted \%), and mean absolute deviation (MAD, percentage points) from the default weight. The inferred income/employment scheme and general-population raking are shown for comparison and are not used for headline results.}
\label{tab:weights}
\small
\begin{tabular}{lccccccc}
\toprule
Scheme & Deff & Eff.\ $n$ & ChatGPT & Gemini & Claude & Concern & MAD \\
 & & & prim. & prim. & prim. & ($\geq$4) & \\
\midrule
Unweighted & --- & 1{,}999 & 59.6 & 24.6 & 7.0 & 80.3 & 0.56 \\
Age\,+\,education (default) & 1.13 & $\approx$1{,}800 & 58.2 & 25.4 & 7.0 & 80.4 & --- \\
Age\,+\,education\,+\,industry & 1.54 & $\approx$1{,}320 & 58.0 & 25.2 & 6.4 & 78.3 & 0.74 \\
\,+\,inferred income/employment & 2.24 & $\approx$890 & 55.5 & 27.3 & 7.8 & 79.5 & 1.58 \\
General-population raked & 3.20 & $\approx$620 & 52.0 & 30.5 & 6.2 & 78.3 & 3.56 \\
\bottomrule
\end{tabular}
\end{table}

The inferred income/employment scheme deserves a brief note since it is the one methodological path we developed but did not take. It infers adoption gradients for income and employment---dimensions not directly measured by our external benchmarks---by pushing the age--education adoption surface through the Census joint distribution under the assumption that adoption is conditionally independent of income and employment given age and education. That assumption is strong and not testable with our data, and because the resulting weights change headline estimates by more than the measured-gradient schemes while resting on a modeling assumption, we report them here only as a robustness bound rather than adopting them.

\subsection*{A.2\quad Engagement by age}
Table~\ref{tab:age} reports weighted engagement metrics by age band with 95\% intervals. Awareness breadth is flat ($\approx$6 of 8 platforms known at every age); functional reliance and paid conversion peak in the 25--44 core; informational use rises with age; trust shows a mild U-shape. Small older cells (55--64, 65+) carry wide intervals and should be read as indicative.

\begin{table}[h]
\centering
\caption{Weighted engagement by age band ($\pm$ = 95\% CI half-width). ``Platforms used'' counts distinct platforms named as primary across task categories.}
\label{tab:age}
\small
\begin{tabular}{lccccccc}
\toprule
Age & $n$ & Aware & Used & \% inform.\ & \% compan.\ & \% pay & \% trust \\
\midrule
18--24 & 265 & $5.84\pm0.23$ & $1.53\pm0.09$ & $81\pm5$ & $34\pm6$ & $35\pm6$ & $35\pm6$ \\
25--34 & 621 & $5.84\pm0.15$ & $1.60\pm0.07$ & $84\pm3$ & $36\pm4$ & $34\pm4$ & $30\pm4$ \\
35--44 & 559 & $5.93\pm0.16$ & $1.61\pm0.07$ & $88\pm3$ & $38\pm4$ & $39\pm4$ & $34\pm4$ \\
45--54 & 319 & $6.14\pm0.20$ & $1.54\pm0.09$ & $89\pm3$ & $33\pm5$ & $28\pm5$ & $30\pm5$ \\
55--64 & 163 & $5.85\pm0.30$ & $1.50\pm0.12$ & $89\pm5$ & $36\pm7$ & $32\pm7$ & $39\pm8$ \\
65+ & 72 & $5.74\pm0.51$ & $1.38\pm0.17$ & $84\pm9$ & $28\pm11$ & $24\pm10$ & $39\pm11$ \\
\bottomrule
\end{tabular}
\end{table}

\subsection*{A.3\quad Opt-out reasons by task}
\begin{figure}[h]
\centering
\includegraphics[width=0.82\textwidth]{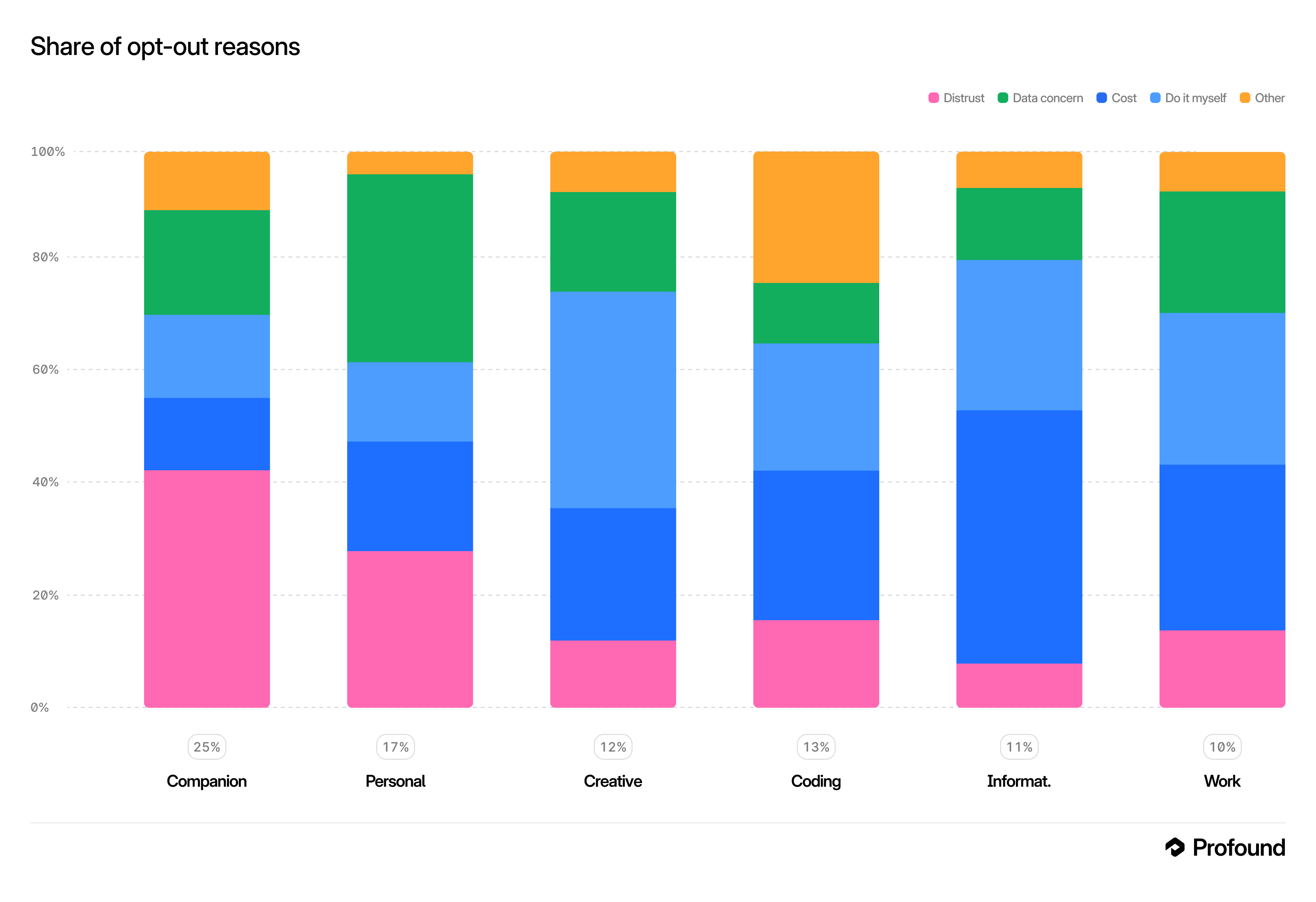}
\caption{Reasons for declining AI by task (share of each task's opt-outs); tasks ordered by opt-out rate (in parentheses).}
\label{fig:optout}
\end{figure}

Figure~\ref{fig:optout} decomposes, for each task, the reasons respondents give for declining AI. Distrust dominates for companionship, data-use concern for personal advice, do-it-myself preference for creative work, and cost for informational, coding, and work tasks.

\end{document}